# The Taming of Plutonium: Pu Metallurgy and the Manhattan Project

Joseph C. Martz, Franz J. Freibert, and David L. Clark
Los Alamos National Laboratory
Los Alamos, NM 87545

**Abstract**: We describe the wartime challenges associated with the rapid developments in plutonium chemistry and metallurgy that were necessary to produce the core of the Trinity Device. Beginning with microgram quantities of plutonium metal late in 1943, initial measurements showed a wide and confusing variance in density and other properties. These confusing results were the first clues to the astounding complexity of plutonium. As this complexity was revealed, it introduced new challenges for the fabrication of kilogram-scale parts. In a remarkable period from January 1944 to June 1945, Manhattan Project scientists made rapid progress in understanding plutonium chemistry and metallurgy. By early 1945, they had discovered five of the six ambient-pressure phases of unalloyed plutonium and reported the density of these phases to within a value of 0.1 g/cm$^3$ of those accepted today. They solved the stability problem introduced by these phases with a rapid alloy development program that ultimately identified gallium as the preferred element to stabilize the delta-phase, producing a plutonium alloy still of scientific and technical interest today. We conclude with a description of post-war developments in these areas, including applications of wartime plutonium metallurgy to civilian applications in nuclear reactors. We dedicate this paper to the memory of Ed Hammel, the Manhattan Project plutonium metallurgist whose previous description and documentation of plutonium history during the war has been essential in our research.

## Introduction

Among the great challenges of the Manhattan Project was enabling the practical use of plutonium as a fission fuel for the atomic bomb. The nuclear properties of plutonium had been measured soon after its discovery and these properties showed that $^{239}$Pu was an excellent candidate for the fissile fuel for an atomic weapon. However, everything else about plutonium and its use in this application was highly uncertain.

The most basic properties of plutonium needed to inform a design of a functioning weapon were unknown beyond its half-life and a crude estimate of its nuclear cross section.[1,2] A functioning design would require knowledge of its density and compressibility.[3,4] Construction of a given design would require limits on impurities and the ability to construct a geometrically-stable shape that would withstand the rigors of assembly, transport, and deployment.[5] Given that only micrograms of material were available when it was decided to pursue a plutonium-fueled weapon, the need to elucidate all of these properties and determine practical means to produce kilogram-scale parts were key problems to be solved as part of the Manhattan Project.

The mysteries of the new element plutonium were waiting to be uncovered.[6–9] The incredible, confounding complexity of plutonium—something well recognized today[10–13]—was completely unknown to the pioneers of the Manhattan Project. This complexity manifested23 itself in conflicting data from measurements on the first tiny bits of metallic plutonium and later proved a vexing challenge to be overcome by intuition, rigorous experimentation, and as these pioneers themselves claimed, good old-fashioned luck.[7,9,14,15] Little did they know, plutonium would prove to be the most complex element on the periodic table.

The story of plutonium during the Manhattan Project has been described in several seminal references.[7,8,14–16] Our review draws heavily upon the work of the renowned plutonium metallurgist Ed Hammel. Hammel summarized key elements of this work in separate volumes of *Los Alamos Science*,[6,8] and published a definitive timeline and personal recollections of plutonium metallurgy during the war in his book "Plutonium Metallurgy at Los Alamos: 1943–1945."[7] In addition, historians at Los Alamos reviewed considerable Manhattan Project source material and described previously unreported details of the plutonium metallurgy story in the book "Critical Assembly."[17]

## The Beginning: the Conflict between Chemistry and Metallurgy Roles

The efforts in plutonium chemistry and metallurgy between 1943 and 1945 had a confrontational beginning.[8] The main conflict existed over the division of labor between Manhattan Project laboratory sites. In 1943, as Los Alamos management, infrastructure, and research was being organized, most of the expertise in plutonium chemistry resided at either the University of California Berkeley Laboratory, where plutonium isotopes were discovered and separated by G. Seaborg, et al.,[14] or at the University of Chicago Metallurgical Laboratory or "Met Lab" in Chicago, where E. Fermi and A. Compton, et al.

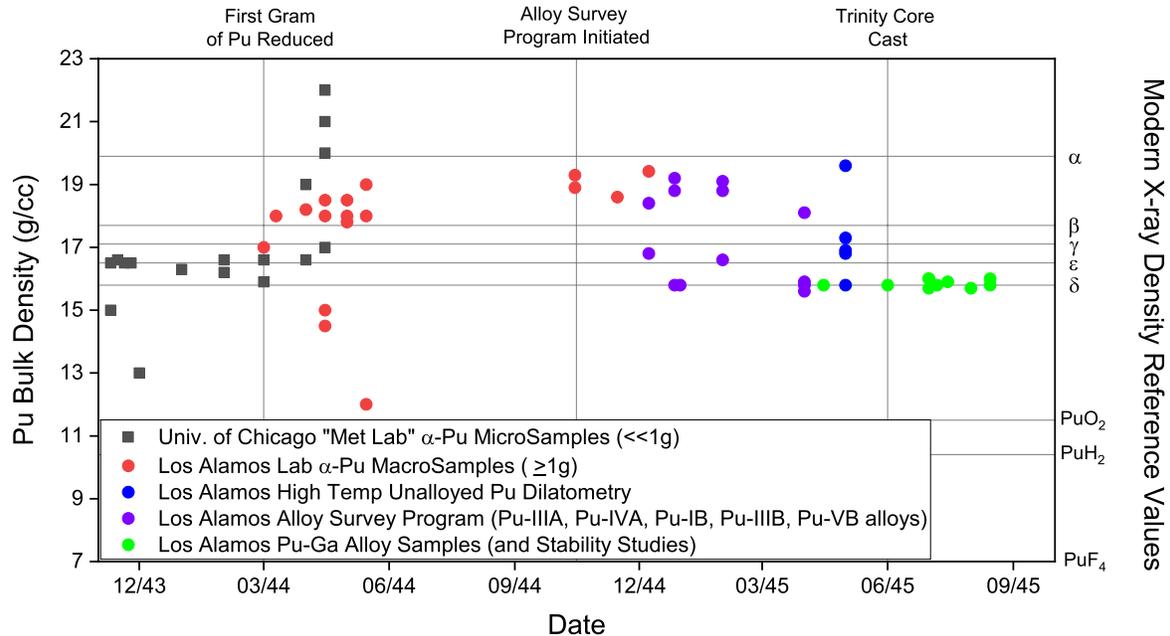

**Figure 1**. A plot of plutonium metal density data showing the variation in measured density (left axis) between December 1943 and September 1945. On the right, we show the modern known densities for important plutonium allotropes and compounds. (color version available online).

were studying uranium metallurgy[18] and operating the CP-1 uranium reactor.[19]

In April 1943, a series of conferences was held at Los Alamos for the purposes of acquainting new staff members with the existing state of knowledge and for preparing a concrete program of research. During these conferences, Los Alamos was visited by the members of a special review committee, appointed by General L. R. Groves and chaired by W. K. Lewis of MIT. The conflict over the Manhattan Project chemistry and metallurgy division of labor was settled in May 1943 when the Lewis Committee recommended that the final purification of plutonium, the reduction to its metallic state, the determination of the metal's relevant physical and metallurgical properties, and the development of the necessary weapon-fabrication technologies would be carried out at Los Alamos. Given that most of the expertise to accomplish these tasks was to be found at either Berkeley or the Met Lab, most of the original Los Alamos chemistry and metallurgy staff came from these two laboratories.[15] It took the remainder of 1943 to arrange the transfer of this staff to Los Alamos.

Meanwhile, preparation and experimental measurements on microgram plutonium samples continued at the Chicago Met Lab. This early experience with micrometallurgy proved invaluable and served to provide the metallurgists at Los Alamos with early data in the critical few weeks before their own gram-scale lots would become available in March 1944.[15] **Figure 1** shows plutonium metal density data obtained in this timeframe.[7]

As more data was collected at both Chicago and Los Alamos in the spring of 1944, the measured density grew in variance. This was among the first clues that plutonium was a complex element, and that many challenges would lie ahead in unravelling these mysteries.

In addition to the density of plutonium, other important chemical and metallurgical properties were identified and became a focus of both the Chicago Met Lab and Los Alamos Laboratory staff. These included:

- chemical analysis (to determine the extent to which purity specifications were being met with the as-reduced metal);
- hardness;
- ductility;
- melting point;
- remelting behavior (including interaction of the molten plutonium with the crucible, gas evolution resulting in spattering, etc.);
- thermal expansion;
- thermal arrests (which, upon heating the sample at a constant rate, provided evidence of phase transitions);
- tensile strength;
- crystal structure determinations; and
- metallographic examinations.[7]

Many of these properties required new and novel means of experimentation. Considerable effort was expended in developing procedures and means to measure these varied properties. These efforts in experimental measurement represent a key, early contribution from all the sites involved with plutonium during the war. The development of new techniques and the resulting physicochemical properties of plutonium metal were described in a 1944 two-volume book edited by Thomas, and Warner,[1] the first comprehensive reference text describing methods and results on this new element.

## Gun Weapon to Implosion Assembly

In July 1944, the Los Alamos Laboratory decided to abandon the development of a plutonium gun weapon due to the high levels of spontaneous fission in Hanford-supplied plutonium.[5,20,21] Focus turned instead to an implosion device. Sufficient plutonium for an implosion design would become available in mid-1945. Given the wartime desire to produce an atomic bomb as soon as feasible, this timeline required an aggressive development effort to meet this challenging schedule. The Los Alamos Laboratory attention had to abruptly refocus from solving the problem of purifying plutonium to developing means to produce precise, stable shapes of plutonium metal for the implosion design.[17]

Recognizing the magnitude of these tasks, the Plutonium Metallurgy Group in Chemistry/Metallurgy (CM) Division expected to utilize the months following July 1944 to measure the physical properties of plutonium metal in a more systematic and orderly way. In addition, this group had numerous additional tasks, including investigating various fabrication techniques, determining the transition rates of plutonium's different allotropic forms, determining the impact of trace impurities, developing scaled-up apparatus to handle larger samples, improving safety practices, and developing techniques for protecting plutonium from oxidation (and its handlers from contamination). By so doing, it was expected that the division would be well prepared to fabricate the plutonium core into whatever shape the weapon designers specified.

Of these multiple goals, the highest priority was the need for a stable and reproducible metallic phase. The growing observation that very small amounts of particular impurities could markedly influence the chemical and/or physical properties of plutonium suggested this task was more complex than first assumed. Further, it was known that these effects were often exacerbated in metals exhibiting several allotropic forms. These metallurgical concerns were documented in an August 4, 1944, memo from C. S. Smith to J. W. Kennedy entitled "Purity Requirements of 49 Metal."[7] Smith started the memo with "*The recent removal of the α→n imposed purity specifications on plutonium must not be construed as meaning that there are no requirements for purity on the material. The specifications on diluting elements (a maximum of 5 atoms per cent) or decrease in density (<0.5 gm/cc) will be surpassed in stringency by strictly metallurgical considerations*." This memo began a

discussion that eventually evolved by October 1944 into the Alloy Survey Program.[7]

In late 1944, Seaborg et al. commissioned a report to summarize what was known of the chemistry and metallurgy of plutonium. The "Chemistry, Purification and Metallurgy of Plutonium, Vol. 1 & 2" edited by C. A. Thomas and J. C. Warner[1] captured the successes and achievements of the Manhattan Project chemical and metallurgical efforts to that date by recognizing the successful production of metallic plutonium within the range of impurity limits for most of the light elements. This remarkable accomplishment was not given to any specific individual or institution, but recognized as the collective body of work by the Ames, Berkeley, Clinton, Chicago, and Los Alamos Projects. Also, general credit was given in a listing of the names of the scientists who had participated in the purification and metallurgy program in the various laboratories.[1] This report can be considered the prelude to the "Plutonium Handbook," first published in 1967 in 2 volumes,[22] and recently updated in 2019 as a seven volume set.[10]

## Purification Challenges

There were many challenges associated with the chemical purification. Plutonium was initially produced in irradiated uranium targets with a maximum concentration of 250–300 ppm. At these low, impurity-like concentrations, plutonium had to be separated from many tons of uranium and fission products and then concentrated. The highly radioactive fission products had to be separated to less than one part in $10^7$ of the original plutonium so that it was safe to handle.[23] Without separation from the fission products, the plutonium from each ton of uranium would have more than $10^5$ Ci of energetic gamma radiation.[24] When the plutonium solutions arrived at Los Alamos, the plutonium had to be chemically converted into high-purity compounds suitable for reduction to plutonium metal. At the beginning of the Manhattan Project when gun assembly was being considered, the original impurity limits specified on the order of a few parts per 10 million by weight of each of the lightest impurity elements![7,8] These almost unheard-of concentrations were necessary to ensure a low neutron background from light atom α-n reactions, that would result in a predetonation or "fizzle" of the proposed device. This required substantial plutonium reprocessing and repurification at Los Alamos.

## Plutonium Metal Preparation

Even before plutonium had become available, metallurgists at Los Alamos and Chicago's Met Lab realized they would need to develop chemical techniques to prepare plutonium compounds that would be suitable for reduction into pure metal.[1,7,25,26] The metallurgists had been working with an exothermic reaction known as metallothermic reduction, in which a plutonium chloride or fluoride compound ($PuCl_3$, $PuF_4$, etc.) is reduced at high temperature in a molten salt containing an active metal like calcium, all sealed inside a metal container referred to as a bomb.[1,25]

Many parameters had to be optimized such as the form of the salt, the reducing agent, their mesh sizes, how reagents were layered in the bomb, the identity and fabrication of the refractory material used as a liner, the heating rate, time spent at maximum temperature, etc. A key concern was how to get pure molten plutonium metal to coalesce so that it could be separated from the slag formed by the reaction products. The problem was exacerbated by the milligram scale required in the early days of the project due to the sheer scarcity of plutonium. Metallurgists worked with uranium as a surrogate and then transferred the techniques to use with plutonium.[7,26] The problem of metal coalescence on a small scale was first solved by T. Magel and N. Dallas working at the Met Lab in Chicago, who moved to Los Alamos to focus on plutonium.[6,25] These workers performed the small-scale reduction and separation inside a graphite centrifuge heated inside an induction furnace. Rotation of the centrifuge at about 50 $g$ was enough to force the molten metal outward to the cone-shaped tip of a refractory liner where it could be isolated as a small button of metal. This ingenious apparatus is illustrated in **Figures 2–4**, with description in the figure captions.[25]

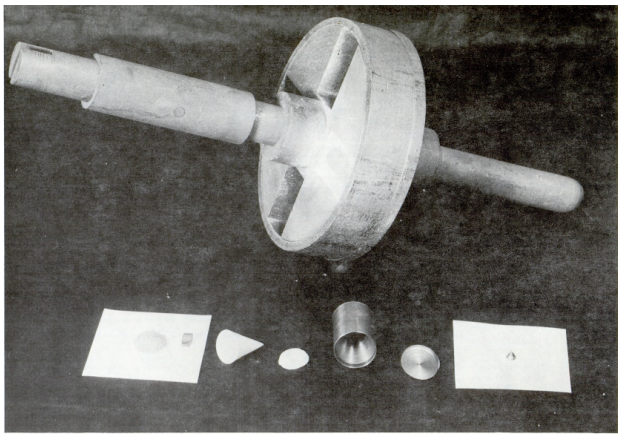

**Figure 2**. Bomb assembly and rotor for the Magel-Dallas Hot Centrifuge reduction. The charge of metal halide ($PuF_4$) and a reducing agent are shown on the paper at the left. This charge was placed into the cone shaped crucible of BeO, with the halide on top, covering the reducing metal (Ca or Li). The crucible was covered with a double lid (at right of the crucible) of a sintered salt (NaCl, LiF, $BaCl_2$) followed by MgO. The crucible was placed inside the cone-shaped interior of the steel bomb that was then sealed by arc welding. The centrifuge contained four slots such that four reductions could be carried out simultaneously.[6,25]

On March 2, 1944, Magel and Dallas produced a 20 mg button; that was the first sample of plutonium metal.

prepared in sufficient quantity to see without magnification.[7] The team produced their first 520-milligram (gram-scale) button of plutonium metal on March 23, 1944.[25] These first metallic samples were crucial to determination of the fundamental properties of plutonium necessary to fabricate into a weapon.

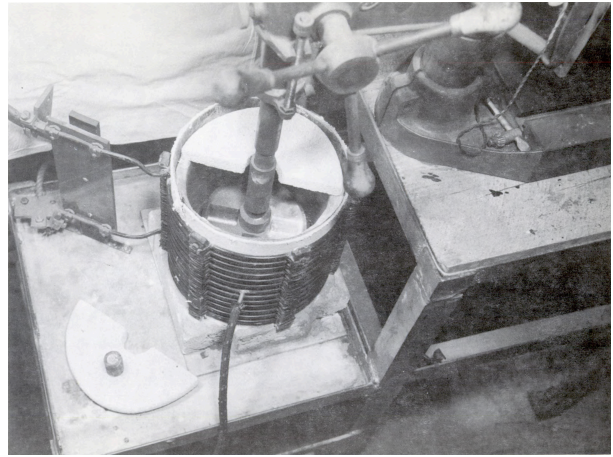

**Figure 3**. Photograph of the centrifuge inside an induction coil for heating. The shaft of the rotor was attached to a drill press so that the apparatus could be rotated at 900 rpm (about 50 $g$) while heating the spinning rotor and bombs to about 1100°C for 3–5 minutes. Rotation was maintained until the apparatus cooled to 400°C –500°C.[6,25]

Working in parallel, R. D. Baker was leading an effort to achieve the same reduction in a stationary bomb (without centrifugation), which was not only more convenient, but essential for the ability to scale-up to the large quantities of plutonium needed for the Trinity Device. Baker's team performed exhaustive studies to identify the most suitable reagents and reaction conditions to continue to scale the process in a stationary bomb.[27–31] They succeeded in performing many large scale reductions, and during the month of May of 1945, they prepared 3.33 kg of metal (page 65).[7] The techniques developed by Baker's team are essentially those used today.[32]

## Plutonium Allotropes

A rigorous scientific study of the properties of this new element was limited by the time constraints and pressure to produce practical information to inform device fabrication. Chemical purity was the initial focus of these efforts. The initial impurity requirements in plutonium were harder to meet than in the case of uranium, which had been employed as a testing and manufacturing process development surrogate. According to the original purity requirements, all operations would have to be carried out in such a way as to avoid contamination with light elements of even of a few parts per million. This requirement resulted in the development of heavy-element refractories for chemical and metallurgical processing equipment (see "Plutonium Crucibles", below).

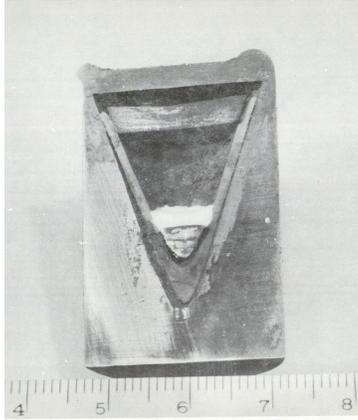

**Figure 4.** A longitudinal cross section of a bomb that was fired in the graphite centrifuge. In this sample, the layer of slag is clearly visible on the top of a button of uranium metal. The button is located in the tip of the crucible.[6,25]

The substantial relaxation of purity requirements that came with the abandonment of the plutonium gun program in mid-July 1944 was welcomed. By this time, the original high-purity goals had nearly been reached. This valuable experience meant that production of sufficiently pure kilogram quantities of plutonium was almost certainly achievable and some steps in the process chain could benefit from relaxed requirements. The attention of CM Division turned to the production of a stable metallurgical phase.

With only microgram quantities of plutonium available, wildly erratic properties were measured on different samples. For example, metal specimens subjected to similar treatments gave dramatically different densities (see **Figure 1**). Measured densities as low as 13.0 g/cm³ were seen in x-ray measurements in late 1943, while values over 22 g/cm³ were later determined at the Chicago Met Lab in the spring of 1944. As larger samples became available in the spring of 1944, density measurements began to converge on two values: many samples, particularly those prepared by centrifuge reduction at Los Alamos, exhibited densities near 15.8 g/cm³ when measured by x-ray diffraction (which also revealed a cubic structure) and with companion immersion densities measured near 14 g/cm³.

Meanwhile, samples prepared at the Met Lab and larger samples prepared by static bomb reduction at Los Alamos showed x-ray densities between 20–21 g/cm³ with a complex, undetermined crystal structure and immersion densities between 18–19 g/cm³. The last centrifuge-prepared sample, the largest produced at the time, also showed these higher densities. Mechanical properties of these samples varied widely as well, with high-density samples reported to be hard and brittle, while the low-density samples were described as soft and malleable.

Of course, today we recognize this variability as the complex phase behavior of pure plutonium along with the role of impurities in stabilizing particular phases, but this fact was unknown to those perplexed by the conflicting data amassed in 1944.[25] With the delivery of the first gram-scale samples in late March 1944, thermochemical measurements began to reveal a series of anomalous properties in plutonium, including more density variation, thermal expansion behavior, and the hint of "arrests" in expansion on heating. By May 1944, M. Kolodney had unambiguously demonstrated that the melting point was below 660°C (page 52)[8], while in June 1944, F. Schnettler "unambiguously demonstrated that transformations into at least two different allotropic forms were associated with the progressive heating of plutonium and that the transition temperature was in the range of 130°C–140°C."[8,33,34] Within weeks, two additional phases had been identified in dilatometer data. A summary of the knowledge in late 1944 is given in **Table 1**.

**Table 1.** As late as December 1944, the knowledge of plutonium allotropes and their properties was incomplete. This table originates from the 1944 Thomas and Warner report entitled "The Chemistry, Purification and Metallurgy of Plutonium"[1]

| Phase | Temperature Range of Existence | Structure | gm/cc at 25°C. | Properties |
|-------|-------------------------------|-----------|----------------|------------|
| α | R.T. – 115°C. | Orthorhombic | 19.8 | Hard, rather brittle |
| β | 115° – 225°C.? | Unknown | 17.8 | Medium hardness, malleable |
| γ | 225°?– 310°C. | Unknown | ? | Very soft and mallable |
| δ | 310° – 485°C. | Face-centered cubic | 16.0 | Very soft |
| ε | 485°?– ?? | Unknown | ? | ? |

By early 1945, Martin and Selmanoff had constructed a high-temperature quartz dilatometer capable of uniquely determining the anomalous length changes that accompany plutonium allotropic transformations and determining the density and thermal expansion of the various phases.[35] Their work clearly showed the presence of five phases in pure, unalloyed plutonium as seen in **Figure 5**. The bulk density of α-phase was reported as 19.6 g/cm³ for temperatures below 130°C, β-phase was reported as 17.3 g/cm³ for averaged temperature range of 150°C–210°C, γ-phase was reported as 16.8 g/cm³ for averaged temperature range of 230°C–300°C, δ-phase was reported as 15.8 g/cm³ over the temperature range of 330°C–470°C. Quite unexpectedly, the δ-phase showed a negative thermal expansion coefficient! Finally, the ε-phase was reported as 16.8 g/cm³ for temperatures above 500°C. Knowing the unique temperature ranges of each allotrope enabled other physical properties to be determined such as crystal structure, electrical resistivity, hardness, thermal conductivity, etc.[36]

In hindsight, it is rather remarkable how accurate these identifications proved to be. In both density and temperature, the values reported in early 1945, using less than a gram of material, were quite accurate. The reported densities are within 0.1 g/cm³ of the accepted values today.[11,37] This fundamental work formed the basis of what has been an active field of research and accumulation of knowledge for the last 75 yrs.[37] In 1945, this information was put to immediate use in determining practical methods to prepare large samples for both experimentation (100-g scale), and ultimately parts for the "Gadget" and subsequent assemblies (kilogram scale). Although five of the six allotropes of plutonium had been discovered, this insight only complicated the fabrication program. Many practical problems remained, including identification of satisfactory materials for processing of molten plutonium as well as the need to produce dimensionally-stable parts from a still-undetermined phase or possibly alloy of plutonium. Note that the sixth recognized allotrope—the delta-prime phase—was finally discovered in 1954,[38] a phase revealed when exceptionally high-purity plutonium was prepared and measured.[39]

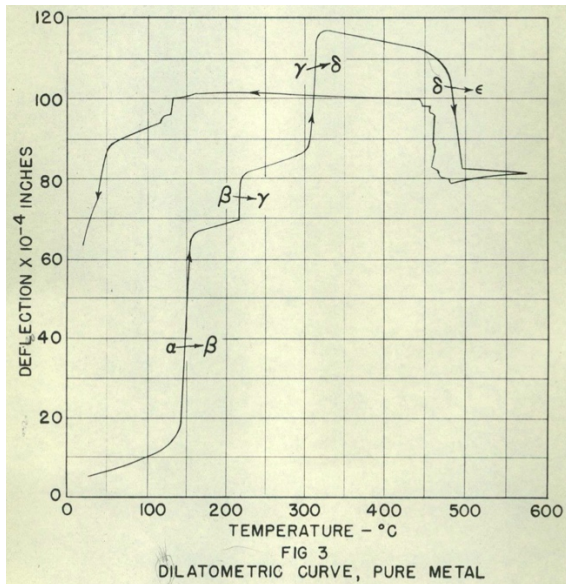

**Figure 5**. Dilatometry trace from 1945 showing the presence of five allotropic phases for pure plutonium.[35]

## Plutonium Crucibles

Planning for the use of plutonium had assumed the properties would most closely align with those of uranium metal. The most important of these were melting point. Uranium melts at the fairly high temperature of 1132°C. A large number of practical considerations are driven by this temperature, including process equipment, casting molds, and most-significantly, crucible materials.

At temperatures above 1000 °C, many common crucible materials such as magnesia were unsuitable. A rather significant effort was undertaken at Berkeley by Wendell Latimer and supervised by E. D. Eastman. This work was later transferred to MIT to develop alternative crucible materials in recognition of this problem.[6,17] Not only was high-temperature structural integrity required, the chosen material must also resist dissolution by molten plutonium, a rather aggressive solvent! This fact, combined with the expected purity requirements that severely limited the acceptable concentration of low-Z elements (the "impurity problem"), required substantial efforts to identify an appropriate crucible material. The solubility of many elements in molten plutonium ruled out a number of common crucible materials, most notably graphite (elemental carbon). More exotic ceramics would be required. **Table 2** provides a summary of crucible research undertaken at Berkeley and MIT, with an emphasis on resultant impurity contamination of the metallic plutonium. Of particular note is the observation that magnesia (MgO) crucibles "reacted vigorously with molten metal at 1000°C" and were obviously unacceptable for high-temperature processing.

Cerium sulfide proved to be the most acceptable candidate. This rather exotic material was supplied by E. D. Eastman's group at Berkeley and codenamed "brass,"[7] and substantial effort was expended at MIT in scaling cerium sulfide production to support kilogram-scale plutonium operations.

**Table 2**. Summary of plutonium crucible research from Berkeley and MIT, 1944 (page 38).[7]

| Crucible Material | Mfg st | Be | Mg | Ca | S | Fe | Ce | Th | Comments |
|---|---|---|---|---|---|---|---|---|---|
| \multicolumn{10}{c}{Impurities in Parts Per Million (by Weight) in Plutonium Remelted in the Crucibles Listed} | | | | | | | | | |
| BeO | Y | ~30 | <5 | ≤5 | --- | 180 | --- | --- | OK if 30 ppm Be tolerable |
| MgO | MIT | <0.3 | 32 | --- | --- | 225 | --- | --- | Reacted vigorously with molten metal @ 1000°C. |
| CeS | Berkeley | <0.4 | 8 | ≤10 | 50 | 100 | ~400 | <2000 | Least contamination of metal from crucible material. |
| Ce₂S₄ | Berkeley | <0.3 | 6 | 12 | 700 | 200 | ~1200 | --- | 49 reduced the sulfide |
| ThS | Berkeley | <0.4 | ≤4 | 7 | 700 | 516 | 360 | 6000 | |
| Th₂S₃ | Berkeley | <0.4 | 8 | 8 | 300 | 85 | <20 | --- | Brown in color |
| Th₄S₇ | Berkeley | <0.4 | 5 | 14 | 900 | --- | <24 | ~2000 | |
| Th₃S₁₂ | Berkeley | <0.4 | ≤3 | 6 | 400 | 175 | <30 | ~4000 | Black in color |
| MgO | MIT | --- | --- | --- | --- | --- | --- | --- | MgO was bonded with 1% Ca₃(PO₄)₂ (crucible reacted violently with Pu) |
| TiNᵃ | MIT | --- | --- | --- | --- | --- | --- | --- | 17.4% porosity |
| TiN + 5% TaNᵃ | MIT | --- | --- | --- | --- | --- | --- | --- | 50.4% porosity |
| TiN + 5% Ce₂S₃ᵃ | MIT | --- | --- | --- | --- | --- | --- | --- | 39.8% porosity |
| TiN + 20% Ce₂S₃ᵃ | MIT | --- | --- | --- | --- | --- | --- | --- | 29.7% porosity |
| CeO₂ + 40% MgO (bonded width resin) | MIT | --- | --- | --- | --- | --- | --- | --- | Too difficult to degas |
| MgO | MIT | --- | --- | --- | >200 | --- | --- | --- | Calcined @ 1300°C, crucible fired @ 2190°C. |

ᵃPorosity too high; too much Pu absorbed by crucible.

When the anomalously low plutonium melting point near 650°C was revealed in May 1944, it greatly simplified a number of the preparations for kilogram scale processing. At this rather low temperature, magnesia was now an ideal crucible material. The difficult work of producing cerium sulfide crucibles undertaken at MIT was no longer necessary. A rather apologetic memo from C.S. Smith to John Chipman at MIT noted that this turn of events "*must be very painful*" for the MIT staff that had spent so much time, effort, and expense on developing these CeS crucibles (page 329).[17] It should be noted, common crucible materials for plutonium today include not only magnesia, but tantalum and graphite! It was later found that coating graphite with materials such as erbium oxide ($Er_2O_3$) was sufficient to prevent crucible erosion in the melt.

**Macro-scale Plutonium Metal Production**

Unusual variations in key physical properties continued to be measured through the summer and fall of 1944. It was not until December 1944 that the α phase was recognized as having a density above 19 g/cm³ with a still unidentified, orthorhombic crystal structure (page 285).[17] Uncertainty in the identification of other plutonium phases continued through late 1944.[17] The brittle nature of the α phase was also recognized, and attempts to fabricate foils and sheets resulted in a recognition that processing in the more ductile δ- or γ-phases might allow successful preparation. These efforts were partially successful in December 1944.[17]

By January 1945, the basic issues relating to chemical processing and reduction of plutonium were in a relatively satisfactory state due to the work of R. D. Baker and his team.[27–31] However, for the fabrication groups the situation was far less optimistic. Concerns remained because the shape of the "active material" (a term often used to describe plutonium) to be used in the weapon had not yet been agreed upon by the weapon designers and partly because many fabrication-related questions still remained to be answered.

On January 2, 1945, the first 160-g batch of material was received from Hanford. This material was first purified via wet chemistry with 95% yield, and then converted to $PuF_4$ with 97% yield. Stationary bomb reduction produced the first 100-g scale plutonium metal samples.[27] "The resulting material was then remelted, reduced, and hot-pressed into two 0.90-inch-diameter hemispheres weighing 60 g each. The researchers concluded that the purity of the metal was 'very good' by current standards" (page 469).[17] Maddeningly, these metal parts saw additional warpage and cracking after the metal stood a day or so at room temperature.[40]

Although the weapon designers were still assuming that the core of the final "gadget" would consist of high-density α-phase plutonium, a real concern existed within the CM groups that it might not be possible to guarantee that any nominally α-phase specimen would retain its finished form indefinitely. Pure metal had to be pressed around 250°C in the γ- or δ-phase field of stability, as neither α nor β metal was malleable enough for processing. A contraction from δ→γ or γ→β, was followed by the (potentially delayed) β→α transformation when each pressing cooled to room temperature. The final stage contraction of β→α resulted in a greater than 11% increase in density (and a corresponding decrease in volume), producing warpage and large cracks, often a day or two after cooling to ambient temperature.[40]

This component fabrication challenge required a bold solution, a solution that came from centuries of metallurgical development: the addition of small amounts of other elements to create an alloy. The key question became, what elements? By late 1944, it was determined that if the δ-phase could be stabilized by alloying, it would undergo no transformation on cooling. This would have obvious advantages for component fabrication.

**Plutonium Alloy Survey Program**

The varied density results from 1943 and early 1944 made it clear that some impurity or impurities occasionally present were stabilizing some soft, low-density form at room temperature. In a memorandum sent to J. W. Kennedy on August 8, 1944, C. S. Smith casually mentioned that consideration be given to the "*addition of a small amount of some impurity to a plutonium melt with the hope that such an addition might retard the transformation to the α-phase.*" Smith subsequently noted that, if such a stratagem were found to work, one would, of course, be always dealing with a metastable state but that, nevertheless, it might be worth pursuing (page 53).[8]

It was soon learned from detailed x-ray diffraction data that the δ-phase had a desirable, cubic crystal structure (specifically face-centered cubic [fcc]) and was normally stable in pure, unalloyed plutonium in the temperature range 300°C–470°C.[35] The attractiveness of this cubic phase to solve the fabrication challenges presented by the brittle α-phase would fuel an active search for one or more elements which, when alloyed with plutonium, would stabilize the δ-phase at room temperature. Despite C. S. Smith's suggestion in August, it was not until late October 1944 that a true alloy-survey program would begin to identify a means to prevent transformation to the brittle α-phase. There were additional drivers for the development of a plutonium alloy. Most importantly, the physicists had requested a lower density plutonium for certain nuclear experiments.[41] The Alloy Survey Program would fulfill this request.

It was recognized at the outset that two types of stability of the δ-phase at room temperature were possible:

**Equilibrium Stability**—the thermodynamically stable phase at room temperature with no possible further transformation; and

**Kinetic Stability or Metastability**—stability in which rates of transformation were so slowed by the alloying elements that the δ phase could be super-cooled to room temperature and retained for weeks or longer without transforming to a phase of higher density.

Metallurgists and other technical staff had hoped that equilibrium stability would be established, but it was recognized that metastability was more likely in view of the early results in which the δ-phase was partially stabilized by impurities. This fact of the metastability of plutonium alloys remains a disputed topic even today.[11,42,43]

With scarce plutonium samples in late 1944, it was not possible to investigate each element independently. Instead, alloys were made that contained as many as five intentionally added alloying elements. Combinations of elements thought unlikely to interfere with each other were selected for initial investigation.

The as-cast densities were determined for all the sample alloys. These alloys were then annealed in vacuum at 390°C, cooled rapidly, and their densities redetermined. Alloys in which both the as-cast and annealed densities were greater than 18.0 $g/cm^3$ were withdrawn from further study. Alloys that exhibited as-cast densities below 18.0 $g/cm^3$ and that showed even lower annealed densities became candidates for further examination and study.

**Table 3** provides a summary of the alloys investigated.[7] At the inception of the survey, a large number of elements were tested in December 1944. As Ed Hammel states:

*"an attempt may have been made to rationalize in some way the initial choices, but it is this writer's opinion that the elements actually used probably depended upon which ones were available at the time in Chem Stock. Nevertheless, a surprisingly good selection, from the point of view of systematically sampling the periodic table, was achieved"* (page 40).[7]

Of the large number of alloys examined, only those containing Al and the combined alloy of Pb, Sn, Ge, and Si showed promising results. Each of these would be tested separately, starting with Si. In January 1945, alloys of Si, Ta, Cu+Ag+Au, and La+Ce+W were prepared, all at 1 at.% concentration. Only Si produced a promising result similar to the Al work from December 1944.

Although a number of potential alloying agents were suggested, almost from the beginning the use of aluminum was guided by a knowledge of its valence. As Ed Hammel stated, "*it should also be noted that the valence ranges exhibited by plutonium, namely 3,4, 5, and 6, overlap those of aluminum and gallium, and this suggests that the chemical bonding properties, and hence the electronegativities, of all three elements are roughly similar*" (page 149).[7] Indeed, this use of aluminum to

stabilize the δ-phase was described by Ed Hammel as "*incredibly lucky!*" Hammel expanded on this point:

*"Several times, in the main text of this report, it has been remarked that, in being forced by time constraints to choose among various technical options, 'we were incredibly lucky' in the choices made.*

*a) Discovering that a dilute aluminum alloy of plutonium stabilized the δ-phase upon cooling to room temperature, and*

*b) after learning that aluminum was an unsuitable alloying additive (for nuclear reasons), we shifted to gallium (for what seemed to us at the time to be a good physicochemical reason) and it worked.*

*Again we were unbelievably lucky, because it was subsequently shown that our argument for so doing, though defensible at the time, can now be accurately characterized by the well-known quotation: "it ain't necessarily so!"* (page 147).[7]

The promising aluminum results in the initial survey led to additional studies in late December 1944 by F. J. Schnettler and A. E. Martin that showed that at least 2 at.% Al alloys were required to stabilize the delta phase to room temperature (see **Table 4**).

In February 1945, investigations continued into Si at concentrations of 2.2, 3.3, and 4 at.%, while additional Al alloys were produced at concentrations of 2, 3, and 4 at.%. (**Table 3**). Results for both Si and Al alloys continued to show encouraging results.[7] On March 9, 1945, the official impurity specifications were presented to CM Division. Unfortunately, these specifications precluded the use of a few percent Al to stabilize the delta phase (page. 54).[8] In response to this limit, a combined 0.5% Al+1% Si alloy was produced that showed good evidence of stabilization (**Table 3**). The desire to find a high-Z element to stabilize the delta phase was attempted with the addition of 2 at.% Ge, which showed no evidence of stabilization. By April 1945, studies expanded to include additional Al+Si alloys and the first attempt at Ga stabilization, with 1- and 3 at.% Ga. Finally, to add to the growing data on Pu alloy systems, U was studied at high concentrations of 8-, 15-, and 25 at.%.

**Table 3.** Plutonium alloys density data from the Alloy Survey Program.[7] Note that gallium was not studied before April 1945.

| Nominal Composition (at.%) | Density (g/cm³) As Cast | Density (g/cm³) After Anneal | Temp |
|---|---|---|---|
| **December 1944** | | | |
| 1 Al | 16.93 | 16.79 | 390°C |
| 0.3 Be | -- | -- | |
| 1 C | 19.1 | 19.1 | 390°C |
| 1 Fe | 18.4 | 18.4 | 390°C |
| 1 Li | 18.4 | 18.8 | 390°C |
| 1 Mn | 18.5 | 18.7 | 390°C |
| 1 Th | 18.5 | 18.7 | 390°C |
| 0.5 U | 18.65 | 18.50 | 390°C |
| 1 Co,1 Ni | 18.1 | 18.4 | 390°C |
| 1 Pt,1 Ir | 18.9 | 18.95 | 390°C |
| 1 Mg,1 Cd,1Zn | 18.74 | 18.8 | 390°C |
| 1 Pd,1 Sn,1 Ge,1Si | 16.5 | 16.9 | 390°C |
| 1 Pd,1 Sn,1 Ge,1Si | 16.5 | 16.3 | 390°C |
| 1 Pd,1 Sn,1 Ge,1Si | 16.5 | 17.0 | 390°C† |
| 1 Ti,1V,1Cr,1Mo,1Zr | 18.7 | 18.7 | 390°C |
| 1 Fe,1 Co, 1 Ni | 18.26 | -- | 390°C |
| **January 1945** | | | |
| 1 Si | 16.8 | 16.6 | 390°C |
| 1Cu, 1Ag, 1Au | 18.9 | 18.8 | 390°C |
| 1La, 1Ce, 1W | 18.8 | 18.8 | 390°C |
| 1 Ta | 19.2 | 19.1 | 390°C |
| **February 1945** | | | |
| 2 Al | 16.47 | 15.60 | 400°C |
| 3 Al | 16.18 | 15.51 | 400°C |
| 4 Al | 15.93 | 15.63 | 400°C |
| 2.2 Si | 16.94 | - - - | 400°C |
| 3.3 Si | 15.79 | 15.49 | 400°C |
| 4.0 Si | 15.89 | 15.44 | 400°C |
| **March 1945** | | | |
| 2 Ge | 17.87 | 17.20 | 400°C |
| 0.5 Al + 1.0 Si | 16.12 | 15.77 | 400°C |
| **April 1945** | | | |
| 1.5 Al+ 1.0 Si | 16.1 | 15.8 | 450°C |
| 0.3 Al+ 1.5 Si | 16.7 | 15.6 | 450°C |
| 0.8 Al+ 1.0 Si | 16.8 | 15.9 | 450°C |
| 1.0 Ga | 16.8 | --- | 450°C |
| 3.0 Ga | 15.9 | 15.8 | 450°C |
| 8.0 U | 19.1 | --- | 450°C |
| 15.0 U | 18.0 | 18.0 | 450C |
| 25.0 U | 18.1 | --- | 450°C |
| 0.5 Al + 1.0 Si | 16.12 | 15.77 | 400°C |

† followed by 18 hr 0°C

**Table 4.** Initial Plutonium/aluminum alloy stability study from December 1944.

| | | Aluminum Alloys | | |
|---|---|---|---|---|
| Date | Sample # | Nominal Atomic Percent | Density (g/cm³) As Castᵃ | After Specified Treatment |
| 12/18 | 5498 2Al A | 0.5 | 17.6 | 17.2 | 1-h anneal @ 400°C, rapid cool |
| 12/18 | 5498 2Al A | 0.5 | 17.6 | 17.45 | 41 h @ 0°C |
| 12/27 | 5498 2Al A | 0.5 | 17.6 | 18.91 | maintained at room temperature for 9 days |
| 12/28 | 5498 2Al A | 0.5 | 17.6 | 18.67 | dilatometer run of 12/27 |
| 12/31 | 5498 2Al A | 0.5 | 17.6 | 18.67 | maintained at room temperature for 3 days |
| 12/27 | 5536F | 1.5 | 16.6 | 16.05 | 1-h anneal @ 400°C, rapid cool |
| 12/27 | 5536F | 1.5 | 16.6 | 17.37 | 3-h anneal @ 400°C, rapid cool |
| 12/27 | 5536F | 1.5 | 16.6 | 16.30 | 22 h @ 0°C |
| 12/30 | 5536F | 1.5 | 16.6 | 17.89 | dilatometer run on 12/29 |
| 12/27 | 5536E | 2.0 | 16.0 | 15.80 | 1-h anneal @ 400°C, rapid cool |
| 12/27 | 5536E | 2.0 | 16.0 | 15.80 | 3-h anneal @ 400°C, rapid cool |
| 12/28 | 5536E | 2.0 | 16.0 | 15.98 | 22 h @ 0°C |
| 12/30 | 5536E | 2.0 | 16.0 | 17.84 | dilatometer run on 12/29 |
| 12/31 | 5536E | 2.0 | 16.0 | 17.62 | dilatometer run on 12/30 |

ᵃOr recast.

Ed Hammel provides insight into this critical period and these decisions:[7]

*"At that time, no one involved was paying special attention to the impurity problem because we had been assured that the "old" limits had been increased by about two orders of magnitude and because that problem no longer appeared to be much of a constraint, at least in the initial survey phases of the alloy program. For that reason, encouraged by the results obtained with the first 1-at. %-Al alloy, by February 1945 we had made and were testing a 4-at. % alloy. Then, during the following month, after the revised impurity tolerances did become available, it didn't take someone (who I don't remember who) very long to notice that the new allowed tolerance for Al was 910 ppm by weight, which translates into 0.5 at.%! In consequence, our prime δ-phase stabilization candidate was suddenly removed from further consideration."* - (page 63).[7]

What elemental addition might replace aluminum? The suggestion to use Ga as a substitute was almost immediately made:

*"Fortunately, the periodic table was reasonably familiar to the CM-Division staff, and almost immediately someone (again, I don't remember who, but in this case it could simultaneously have been several staff members) suggested trying the element below Al in that table, namely, gallium. As already noted, this was done in April 1945, and the 3-at. %-Ga alloy turned out to be remarkably stable. Furthermore, with an atomic number more than twice that of aluminum, its (α, n) yield was negligible."* (page 63).[7]

One question that has arisen in our review of the prior literature is: when exactly was Ga first discussed or investigated as a δ-phase stabilizer? In particular, was Ga considered as part of the initial Alloy Survey Program?

Given the importance of Ga as the preferred delta stabilizer, we wanted to confirm its first use as discussed above. We can find no other references to Pu/Ga alloys before April 1945 in the Los Alamos records we reviewed. The only references to Ga we located before this date were its potential use as a "surface rub" to help with forming operations such as rolling.[44]

Rather remarkably, Ga, which would become the preferred δ-stabilizer for plutonium, was suggested and tested only after the initial candidate, Al, was disqualified on the basis of impurity concerns (e.g., the unwanted production of neutrons from (α, n) reactions with light nuclei).

**Plutonium-Gallium Alloys**

Fortunately, 3 at.% Ga was considerably lower in concentration than the acceptable limit for the "impurity problem".[7] Even 1 at.% Ga stabilized the δ-phase alloy to room temperature, though this 1 at.% alloy did transform at –20°C. Alloys with 2 at.% or more Ga did not transform even after being held at –75°C for three days, but their long-term stability was still doubtful. Other measurements on the new Pu-Ga alloys provided further, encouraging data.[41] In another "stroke of luck", the thermal expansion coefficient of the δ-phase was quite low and did not appear to be greatly influenced by the Ga content, a feature that would aid stability in device assembly and transport.[41]

Further investigations in April and May of 1945 showed that the δ-phase was well-stabilized in Ga alloys by annealing at 410°C for 16 hours.[7] After such a "homogenization" treatment, the alloys were stable at any temperature down to –75°C, excepting the 1 at.% Ga alloy that still showed transformation to α-phase at –20°C after homogenization. The 2 at.% and 3 at.% Ga alloys were found to be stable in liquid nitrogen at –195°C. The other alloy compositions were not tested at this extremely low temperature. A series of tests on 18 samples of the 3 at.% Ga alloy was conducted to establish the stability of this alloy during this study.[41]

The pressure stability was examined when 3 at.% Ga alloy was partially converted to α phase by the application of pressures of greater than 100,000 psi (6.9 kbar) at room temperature or at dry ice temperatures of –75°C. Once a δ-phase alloy was partially converted to α-phase by pressure, it could usually be further converted to α by extended low-temperature treatments at –195°C. Such treatments, discovered during the Manhattan Project, would form the basis of metallurgy and phase transformation research of plutonium and its alloys for decades afterward.[13,37,45] These same treatments also demonstrated that the 3 at.% Ga alloy was indeed metastable. With appropriate time, temperature, and/or pressure, it could be converted to an equilibrium mixture of α-phase and various Pu-Ga intermetallic compounds.

In May 1945, the selection of a 3 at.% Ga alloy for the Gadget core represented the end of the Alloy Survey Program. The metastability of this alloy also meant that longer-term monitoring for phase stability would be required. The Alloy Survey Program transitioned to what was called the "Long Time Surveillance Program" for Ga-stabilized, δ-phase plutonium. This research effort was conducted in May and June of 1945.[7] The 3 at.% Ga alloys were kept at temperatures of –75°C, 0°C, 70°C, and 100°C, for a total of 44 days, with no indication of any transformation as determined by density measurements.[7] Once again, "incredible luck" had intervened and the preferred alloy was shown to be stable for 6 weeks, a period of time thought sufficient for production, transport, assembly, and deployment of the atomic bomb.

In hindsight, the progress in plutonium alloy development in 1945 was rather remarkable. The speed with which first Pu-Al, and then Pu-Ga alloys were made and studied is simply stunning. This program had innumerable opportunities for failure. Had the alloys proved insufficiently stable, the entire logistics chain for deployment of the bomb would have been severely impacted at best. At worst, the plutonium bomb would fail not for reasons of physics, but for lack of metallurgical stability. The metallurgists and chemists at Los Alamos made rapid decisions and choose the right materials in the right concentrations to avoid a key problem in the stability of the plutonium core. In so doing, they pioneered the rich field of plutonium metallurgy in a matter of a few months. As post-war plutonium metallurgy expert Fred Schonfeld stated, these developments in 1945 "*probably roughed out more than sixty percent of the field,*" thus making a *"very impressive" beginning to the understanding of the basic characteristics of plutonium metal"* (page 285).[17]

Ed Hammel has attributed much of this work to sheer luck.[6–8] A careful review of their accomplishments and timelines shows they had a deep understanding of alloy theory along with excellent experimentation and measurements. The sum of this data, coupled with their understanding of concepts such as electronegativity and atomic volumes, shows that their "luck" was more the result of excellent preparation and insight intersecting with critical opportunity.

**The Trinity Device and the Legacy of its Plutonium Metallurgy**

Plutonium metallurgy remained an issue of concern for the Los Alamos Laboratory up to the moment of the Trinity Test. Apprehension over plutonium reactivity and corrosion led to the Gadget core design, including an encapsulation layer for passivation.[7] The hot-pressed hemispheres of plutonium that formed the bomb cores were heavily nickel-plated in order to prevent oxidation. The Fat Man bomb core was successfully nickel-coated using nickel carbonyl. However, the hemispheres for the Trinity Device were electroplated and some aqueous electrolyte retained in a porous spot in one of the hemishells reacted and caused a tiny blister to form. This blister was sufficient to separate the mating surfaces enough to allow gas jetting during implosion and possible insufficient compression.

Postponement of the test was threatened, but C. S. Smith proposed the insertion of some rings of crinkled gold foil to prevent jetting and remarkably made the gold foil shims himself.[46] This last-minute fix required his membership on the team responsible for the final assembly of the Trinity Device at the Trinity Test Site. Of the event, C. S Smith recalled in 1981, "*At approximately* *noon on 15 July 1945, at MacDonald's Ranch near Alamogordo in New Mexico, I put the proper amount of gold foil between the two hemispheres of plutonium. My fingers were the last to touch those portentous bits of warm metal. The feeling remains with me to this day, thirty-six years later.*"[15]

Plutonium metallurgy wartime activities continued at Los Alamos Laboratory until September 1945 when nearly all chemical and metallurgical operations in CM Division were shut down in preparation for a major "clean-up" activity that served as a termination point for wartime activities.

## Postwar Plutonium Metallurgy

Advances in plutonium metallurgy continued shortly after the war. The largest mystery involved the structure of the α phase. Zachariasen and Ellinger deconvolved the intricate x-ray data to ultimately determine the complex, mineral-like structure. This work was performed in 1946, but remained unpublished in the open literature until 1957.[47]

Zachariasen and Ellinger[47] demonstrated that the α phase is a monoclinic crystal structure ($P2_1/m$) with 16 atoms per unit cell and eight unique atom positions (see **Figure 6**). As later developments showed, the low-symmetry monoclinic ground-state α phase of plutonium results from the peculiar and non-equivalent nature of the 5*f* electron bonding in plutonium.[48,49]

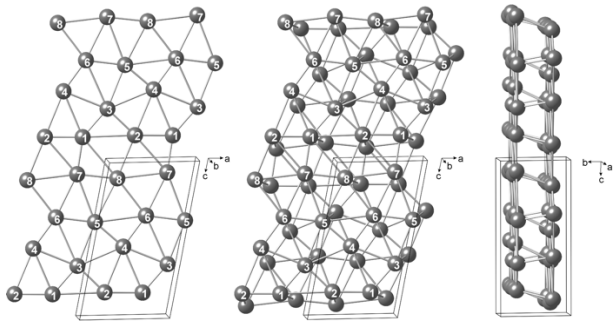

**Figure 6**. Several views of the monoclinic α-phase structure of plutonium with 16 atoms per unit cell and eight different atom positions.[11]

The monoclinic structure of α plutonium is a slight distortion from a hexagonal lattice. This similarity was used by Crocker[50] in modeling the crystallographic relationships of phase transformations involving the α phase. The low-symmetry monoclinic structure of the α phase has a profound influence on its properties; it has no macroscopic ductility, and most properties are highly directional.[51] In many ways, the alpha phase holds more in common with minerals than it does a metal.[52,53] No wonder the Manhattan Project scientists had problems in fabricating macroscopic parts from this enigmatic material!

With the crystallographic discovery of the α-phase structure and the identification of the delta-prime phase, the complete, ambient pressure behavior of unalloyed plutonium was finally resolved in the late 1950s. An idealized thermal expansion curve for pure plutonium is shown in **Figure 7**,[54] along with the known crystal structures and densities of the solid phases and the liquid phase. The stability ranges and crystal structure data for the individual allotropes are shown in **Table 5**.

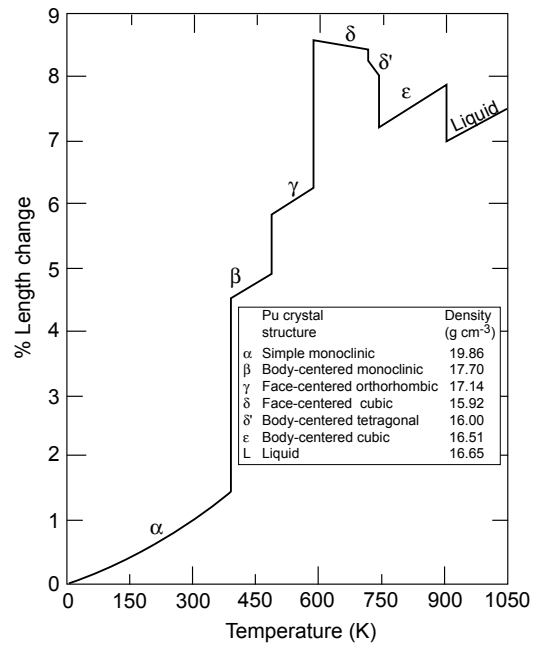

**Figure 7**. Thermal expansion of unalloyed plutonium. This idealized curve was generated by Schonfeld and Tate [55] based on the best available expansion data.

The data shown in **Figure 7** and **Table 5** are the ultimate manifestation of the complexity and instability of plutonium. The six solid allotropes at ambient pressure are the most of any element in the periodic table. This complexity—initially unknown to the investigators during the war—would complicate and confound the myriad tasks needed to ensure the success of the Manhattan Project.

The final discovery in the complex phase behavior of plutonium was made by Morgan in 1970 when he clearly demonstrated the existence of a seventh phase at high pressure, designated ζ phase (**Figure 8**).[56] Even today, the ζ phase structure has yet to be determined.

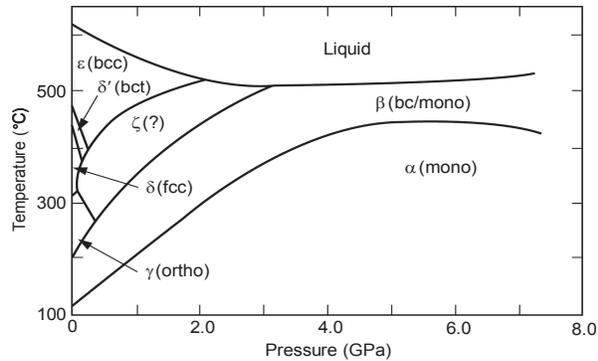

**Figure 8.** Pressure-temperature phase diagram from Morgan showing the existence of a seventh phase, $\zeta$.[56]

**Table 5.** Crystal structure data for plutonium.[54]

| Phase | Stability range (°C) | Crystal lattice & space group | Atoms per unit cell | x-ray dens. (g/cm³) | Trans temp (°C) |
|-------|---------|---------|------|-------|------|
| $\alpha$ | below 122 ± 4 | simple monoclinic P2₁/m | 16 | 19.86 | |
| $\beta$ | (122 ± 4)– (207 ± 5) | body-centered monoclinic I2/m<sup>a</sup> | 34 | 17.70 $\alpha{\rightarrow}\beta$ | 122 ± 4 |
| $\gamma$ | (207 ± 5)– (315 ± 3) | face-centered ortho Fddd | 8 | 17.14 $\beta{\rightarrow}\gamma$ | 207 ± 5 |
| $\delta$ | (315 ± 3)– (457 ± 2) | face-centered cubic Fm3m | 4 | 15.92 $\gamma{\rightarrow}\delta$ | 315 ± 3 |
| $\delta'$ | (457 ± 2)– (479 ± 4) | body-centered tetrag. I4/mmm | 2 | 16.00 $\delta{\rightarrow}\delta'$ | 457 ± 2 |
| $\varepsilon$ | (479 ± 2)– (640 ± 2) | body-centered cubic. Im3m | 2 | 16.51 $\delta'{\rightarrow}\varepsilon$ | 479 ± 4 |
| | | | | m.p. | 640 ± 2 |

Beyond the defense application of plutonium, there was a recognition that plutonium had numerous advantages for civilian nuclear power, advantages enabled by an understanding of plutonium metallurgy. Director N. Bradbury proposed a metallic-fuel reactor concept by letter to Major General L. R. Groves on November 23, 1945, and approval was granted by December 17, 1945, to pursue this concept.[57] Although the reactor required 17 kg of plutonium, the plutonium allotted was not completely satisfactory for weapon use. Ground was broken for the new laboratory building in Los Alamos Canyon in May 1946.

The reactor fuel was intentionally chosen as 3 at.% Ga-stabilized δ-phase plutonium to ensure stability at high temperature. Additional advantages were that delta-phase thermal conductivity was higher than any other phase of plutonium and its metallurgy was considered fairly well known.[58] Experiments on rods of the reactor material began in March 1946 within CM Division. The extrusion of Pu-Ga alloy δ-phase rod at low temperature led to partial transformation to α-phase induced by the great amount of mechanical work during extrusion. High-temperature extrusion, all in the δ-phase stable region, proved unsatisfactory from the point of dimensional stability but the resultant density indicated stable δ-phase components. Machining of such extruded oversize rods was undertaken in April 1946 and production started in June 1946. In September 1946, Phillip Morrison accepted a position at Cornell University and Jane Hall and David B. Hall continued as project leaders. Phillip Morrison gave the Los Alamos Fast Plutonium Reactor the nickname "Clementine" and project staff as "miner 49ers" to avoid the classified topic of the fast reactor and plutonium, respectively, in a telegram inquiring as to the Fast Plutonium Reactor operational status.[59]

The Los Alamos Fast Plutonium Reactor "Clementine" eventually achieved a full operating power level of 25 kW in March 1949 and produced a fast neutron flux of approximately $4 \times 10^{12}$ neutrons/cm²/s. On December 24, 1952, the mercury coolant was found contaminated by alpha particle activity and verified plutonium was free in the cooling system, producing a serious hazard.[60] Further Clementine operations were cancelled and the reactor was disassembled. The specifics as to the failure mechanism of the fuel rod cladding remains undetermined; however, this is the only recorded instance in which a weapon alloy was employed as a reactor fuel (**Figure 9**). During the last year that Clementine was operated, the total neutron cross sections of 41 elements were measured with an accuracy of $\pm 10$ percent over a neutron energy range of 3 to 13 million electron volts.[60] These data were of great utility to theorists engaged at that time in the design of both fission and fusion bombs. Furthermore, the Clementine fuel irradiation history is meaningful when one considers equivalence of acute fission damage to hostile weapons environment and chronic low fission damage to long-term self-irradiation damage and plutonium aging.

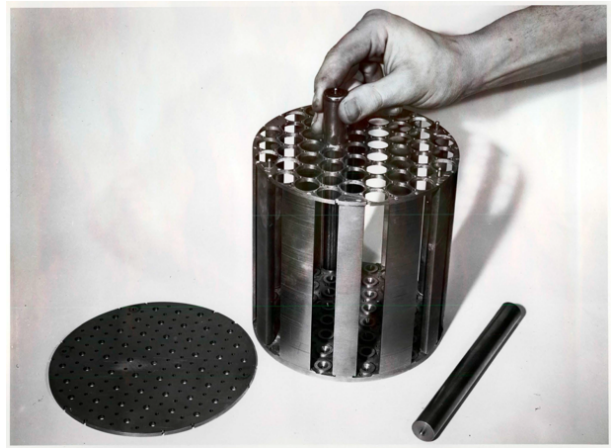

**Figure 9.** Core of Clementine and clad fuel rods.[58]

A final reference to the legacy of the Los Alamos Laboratory plutonium metallurgists are studies that were extended into the liquid state of plutonium metal and some of its alloys circa 1950–1960s. Thermophysical properties of molten fuels were necessary for the fast breeder reactor development activities of LAMPRE (the Los Alamos Molten Plutonium Reactor Experiment).[59] Molten fuels, although metallic in density, do not suffer from the irradiation damage and dimensional instabilities of the solid form, so higher fuel burnup is feasible. Although chemical and metallurgical compatibility between the fuel and cladding at normal reactor operating temperatures is generally not a problem with uranium metal fuels, alloys containing plutonium show undesirable solid-state or molten eutectic reactions with stainless steel cladding materials at expected reactor operating temperatures. Results of these and other studies of plutonium and its alloys have been reviewed recently.[37]

## Acknowledgements

We are grateful to Mark Chadwick, Alan Carr, Sig Hecker, Steve McCready, Scott Crocket, and the archives team at Los Alamos National Laboratory for productive discussions and comments. We dedicate this paper to the memory of Ed Hammel, the esteemed Manhattan Project plutonium metallurgist whose previous work has been essential in our preparation of this manuscript (**Figure 10**). This work was supported by the US Department of Energy through the Los Alamos National Laboratory. Los Alamos National Laboratory is operated by Triad National Security, LLC, for the National Nuclear Security Administration of the US Department of Energy under Contract No. 89233218CNA000001.

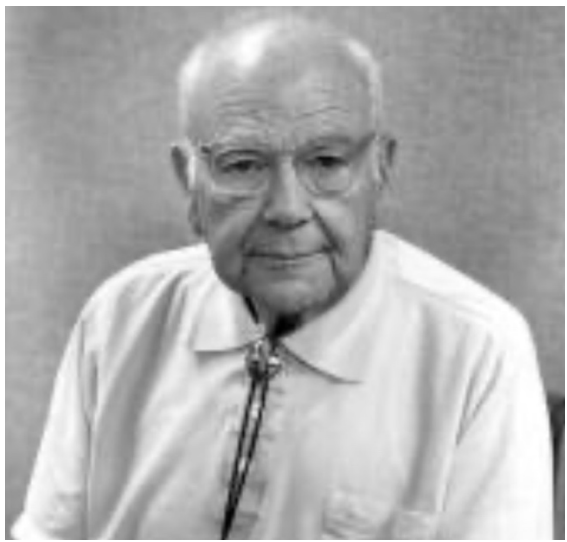

**Figure 10**. Ed Hammel, Ph.D. circa 1995.